\input{aipcheck}

\documentclass[
  ,draft            
  ]
  {aipproc}

\layoutstyle{8x11double}
\def\by#1#2{{\displaystyle {#1}\over \displaystyle {#2}}}
\def\d{{\rm d}}

\begin{document}

\title{Tau contribution and precision measurement of $\theta_{23}$ at
  a neutrino factory}



\author{D. Indumathi}{
  address={The Institute of Mathematical Sciences, Chennai 600 113, India.}
}

\author{Nita Sinha\footnote {presenting author}}{address={The Institute of Mathematical Sciences, Chennai 600 113, India.}}

\begin{abstract}
  We discuss precision measurements of the leading atmospheric
  parameters at a standard neutrino factory. The oscillation of the
  muon and electron neutrinos (anti-neutrinos) to tau neutrinos
  (anti-neutrinos) adds to the muon events sample (both right sign and
  wrong sign) via leptonic decays of the taus produced through
  charge-current interactions in the detector. We focus on how this
  contribution affects a precision measurement of the atmospheric
  mixing parameters and the deviation of $\nu_\mu\leftrightarrow
  \nu_\tau$ mixing from maximality.
\end{abstract}

\maketitle


\section{Introduction}
A neutrino factory, with decays of $\mu^+$ or $\mu^-$ beams in the
straight sections of a storage ring, produces a spectrum of electron
and muon neutrinos via $\mu^+\!\!\to\overline{\nu}_\mu e^+ \nu_e$ and
$\mu^-\!\!\to\nu_\mu e^- \overline{\nu}_e$. Charged-current (CC)
interactions of the muon anti-neutrinos (neutrinos) in a detector lead
to the production of $\mu^+$ ($\mu^-$) or the ``right-sign'' (RS)
muons. The electron neutrinos (anti-neutrinos), on the other hand, can
{\it oscillate} into muon neutrinos (anti-neutrinos) producing muons
with charge opposite to that of the unoscillated case and get detected
as ``wrong sign'' (WS) muons.

Most studies with neutrino factories\cite{IDS-ISS} have focused on the
use of WS events to pin down the unknown reactor angle
$\theta_{13}$, the CP violating phase $\delta_{CP}$ and the mass
hierarchy. RS muons are useful for precision measurements of the
atmospheric mixing angle $\theta_{23}$ and the mass squared
difference $\Delta m^2$, apart from understanding cross-section and
flux uncertainties. In particular, they are sensitive probes of
whether $\nu_\mu\leftrightarrow \nu_\tau$ mixing is maximal (i.e.
$\theta_{23} = \pi/4$, referred to as maximal
$\theta_{23}$). Measurement of deviation from maximality is of
significance in developing models for neutrino masses and mixings.

We address the little-studied issue of contamination of the (RS or WS)
muon events sample from oscillations of the muon or electron neutrinos
(anti-neutrinos) to tau neutrinos (anti-neutrinos), which, through CC
interactions, result in tau leptons that decay to muons. We focus on
how this contribution affects a precision measurement of $\theta_{23}$
and its deviation from maximal mixing.

The oscillation probability $P_{\mu\tau}$ is large even if
$\theta_{13}$ vanishes, being driven by a nearly maximal
$\theta_{23}$. In spite of CC cross-section suppression for the massive
tau production, there is still a sizeable production rate and tau
decay (with rate into muons of $\sim 17\%$) enhances the RS muon
event rates, especially at small muon energies. Since the
$\theta_{23}$ dependent terms in $P_{\mu\mu}$ and $P_{\mu\tau}$ come
with opposite sign, the combination of muons from direct production
and from tau decays marginally {\em decreases} the sensitivity of the
event rates to this angle. Cuts imposed to remove the tau events
drastically reduce the direct muon events as well, worsening the
sensitivity to the oscillation parameters.  Thus, the total muon
events (direct muons and from tau decay) are less
sensitive to the deviation from maximality. Neglect of the tau
contribution will lead to an incorrect conclusion about the precision
possible for the deviation from maximality.
\section{The Inputs: Fluxes, kinematics, cross-sections}

\subsection{The neutrino factory fluxes}

We assume a basic muon storage ring configuration \cite{IDS-NF-base}
with muon beam energy $E_b = 25$ GeV, with $n_\mu = 5 \times 10^{20}$ useful
decays per year. 
We integrate the neutrino fluxes over the muon beam angle $\alpha$,
assuming a gaussian angular divergence of the muon beam around the
$z$-axis with standard deviation \cite{broncano} $\sigma =
0.1/\gamma$, where $\gamma = E_b/m_\mu$. This is then averaged over a
small neutrino opening angle, $\theta' < \epsilon = 0.3\sigma$ or
roughly 0.1 mr. The resulting neutrino (or antineutrino) spectrum with gaussian angular spread averaged
over $(\theta' < \epsilon, 0 \le \phi' \le 2\pi)$ at the base-line
distance $L$ from source to detector is given by, {\small
\begin{eqnarray}
\frac{\d N_\mu}{\d E_\nu}\!\!&\!\!\equiv\!\!&\!\!\frac{1}{E_b}\frac{1}{\int\!\d\Omega} \, \int \d\Omega
\left\langle \by{\d N_\mu}{\d y\d\Omega}
\right\rangle_{\!G}\nonumber \\ 
&\!\!=\!\!&\!\!\by{4n_\mu \gamma^4 y^2}{\pi L^2 E_b} \!\!\left\{\!\! 3
-4 y\gamma^2  -\by{\beta}{2}\left(3-8y\gamma^2\!\!\right)\!\!\left(1+c_\epsilon\right)
e^{-\sigma^2/2}\right. \nonumber \\
& &\!\!\!\!\left.\!\! - \by{1}{3}y\left(\gamma^2\!\!-1\!\!\right)
\left[4+c_\epsilon+c_\epsilon^2+e^{-2\sigma^2}3c_\epsilon
\left(1+c_\epsilon\right)\right]\!\!\right\}, \nonumber \\ 
\frac{\d N_e}{\d E_\nu}\!\! &\!\!\equiv\!\!&\!\!\frac{1}{E_b}\frac{1}{\int\!\d\Omega} \,
\int \d\Omega \left\langle \by{\d N_e}{\d y\d\Omega}
\right\rangle_{\!G} \nonumber \\ 
&\!\!=\!\!&\!\!\by{24n_\mu \gamma^4 y^2}{\pi L^2 E_b}\!\!\left\{\!\! 1
-2 y\gamma^2\!\! -\by{\beta}{2}\left(1-4y\gamma^2\right) \left(1+c_\epsilon\right)
e^{-\sigma^2/2}\right.\nonumber \\
& &\!\!\!\!\!\!\!\!\!\left.\!\!- \by{1}{6}y\left(\gamma^2\!\!-1\!\!\right)
\left[4+c_\epsilon+c_\epsilon^2+e^{-2\sigma^2}3c_\epsilon
\left(1+c_\epsilon\right)\right]\!\!\right\},
\end{eqnarray} 
}
where $y=E_\nu/E_b$, $c_\epsilon,s_\epsilon$ refer to $\cos\epsilon, \sin\epsilon$.

\subsection{The kinematics}\label{sec:kinem}

\begin{figure}
\includegraphics[width=0.4\textwidth]{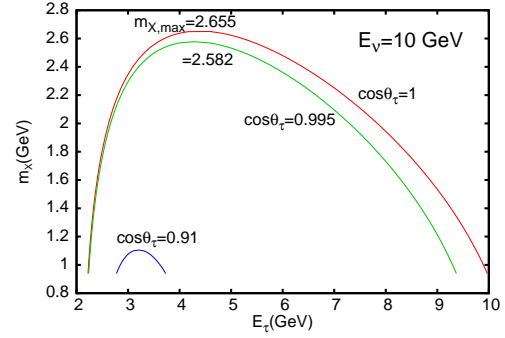}
\caption{Kinematics of $\nu_\tau$--nucleon CC interactions. The
  allowed parabolas of constant $\cos\theta_\tau$ in the
  $m_X$-$E_\tau$ plane are shown for $E_\nu=10$ GeV. The ends of the
  parabolas (at $m_X =M$) give the limits of the tau energy.}
\label{fig:emX}
\end{figure}

We focus on the spectrum of the final state muons and hence require
the detailed kinematics of the CC interactions, in which either muons
or taus are produced, with the latter decaying into muons, where again we use
the differential decay rates (see Ref.~\cite{IS} for details). In the
laboratory frame, a neutrino of flavor $l = \mu$ or $\tau$ interacts
with a nucleon and produces the corresponding charged lepton $l$ at an
angle $\theta_l$ w.r.t. the incident neutrino direction. In the case
of $\nu_\tau$ interactions, the tau is produced at a very forward
angle while the azimuthal angle of the muon from tau decay $\phi_\mu$,
is restricted by the decay kinematics.  The available phase space is
restricted in both direct muon and tau production due to the
constraint on the available energy for the lepton: $E_-< E_l < E_+$;
see Ref.~\cite{IS} for the detailed expressions on the constraints.
The effect of this pinching in available energy, for the case of a
$\tau$ lepton being produced, can be seen in Fig.~\ref{fig:emX} where
the final hadronic mass $m_X$ is plotted as a function of $E_\tau$.
The notation is standard: $m_X^2 = W^2 = (p+q)^2$, where $p$, $q$ are
the nucleon and intermediate gauge boson 4-momenta in the laboratory
frame. For a typical neutrino energy $E_\nu=10$ GeV, allowed energy
range for $\cos\theta_\tau=0.91$ is tiny; tau leptons are hence
produced in a very forward direction while the direct muons, due to
their lighter mass, are less restricted. 

\subsection{The cross-sections}

Since the energies of interest range from a few GeV to 25 GeV, the CC
interactions include quasi-elastic (QE), resonance (Res) and deep
inelastic (DIS) processes. We consider the double differential
cross-sections,
{\small
\begin{equation}
\by{\d\sigma}{\d E_l \d \cos\theta_l} = \by{G_F^2 \kappa^2}{2\pi}
\by{p_l}{M} \left\{ \sum_{i=1}^{5} a_i W_i\right\}~,
\label{eq:cross}
\end{equation}
}
where $G_F$ is the Fermi constant, $\kappa=M_W^2/(Q^2+M_W^2)$ is the
propagator factor with the $W$ boson mass, $M_W$, $p_l$ is the
magnitude of three-momentum of the charged lepton produced and
$W_i$ are structure functions corresponding to the general
decomposition of the hadronic tensor. $W_{4,5}$ appear only
for massive final leptons. We have,
{\small
\begin{eqnarray}
\sum_{i=1}^{5} a_i W_i\!\!\!\!\!\!\!\!& = &\!\!\!\!\!\!\!\!\left(\!\!2W_1+\by{m_l^2}{M^2} W_4\!\!\right)
\left(E_l-p_l\cos\theta_l\right) + W_2 \left(E_l +
p_l\cos\theta_l\right)~ \nonumber \\
& &\!\!\!\! \pm \by{W_3}{M} \left(\!\!E_\nu E_l + p_l^2 -
  \left(E_\nu+E_l\right) p_l\cos\theta_l \!\!\right) - \by{m_l^2}{M} W_5. 
\label{eq:Wi}
\end{eqnarray}
}
The detailed expressions for $W_i$ are taken from Ref.~\cite{yokoya}
where the specific structure functions are listed for QE, Res and
DIS leading order (LO) processes. 
\section{Event rates in a far-detector}\label{sec:event}
\subsection{Preliminaries}\label{sec:prel}

We assume that the neutrinos interact with a 50 kton iron detector
such as the proposed INO/ICAL or MIND. Both $\mu^+$ and $\mu^-$ beams
with equal exposure are considered. While RS events are sensitive to
deviations of $\theta_{23}$ from maximality, inclusion of WS events
may only marginally worsen the results; however, the advantage in
being ``charge-blind'' is significant, hence, all muon events are
simply added.  The generic number of muon events at a distance L, as a
function of the observed muon energy $E$ is, {\small
\begin{eqnarray}
{\cal{R}}^{b}_{i,D}(E)\!\!\!\!\! &=&\!\!\!\!\! K\!\!\int_{E_\nu^{\rm
    thr}}^{E_\nu^{\rm max}}\!\!\!\!\!\!\d E_\nu \by{\d N_i(E_\nu,L)}{\d
  E_\nu}\cdot P_{i\mu}(E_\nu,L) \int_0^1 \,\d\!\cos\theta_\mu \nonumber \\
& & \int_{E_-}^{E_+} \d E_\mu\by{\d\sigma_\mu (E_\nu,E_\mu,\theta_\mu)}
{\d E_\mu\d\cos\theta_\mu} \cdot R(E_\mu,E)~, \nonumber \\
{\cal{R}}^{b}_{i,\tau}(E)\!\!\!\!\! & = &\!\!\!\!\! K\!\!\int_{E_\nu^{\rm thr}}^{E_\nu^{\rm max}}\!\!\!\!\!\d E_\nu  \by{\d N_i(E_\nu,L)}{\d E_\nu}
 \cdot P_{i\tau}(E_\nu,L) \int_{c_{\rm
     min}}^1\!\!\!\!\!\!\!\d\!\cos\theta_\tau\nonumber \\
& &\!\!\!\!\int_{E_-}^{E_+}\d E_\tau
\by{\d\sigma_\tau (E_\nu,E_\tau,\theta_\tau)}{\d E_\tau  \d\cos\theta_\tau} \int_{\rm restr.}\!\!\!\!\!\!\!\!\d E_\mu \d\!\cos\theta_\mu \d\phi_\mu\nonumber \\
& &\hspace{0.5cm}
\by{1}{\Gamma_\tau} \by{\d\Gamma(E_\tau, E_\mu, \theta_\tau, \theta_\mu,\phi_\mu)}{\d E_\mu
\d\cos\theta_\mu\d\phi_\mu}\cdot R(E_\mu,E)~,
\label{eq:rate} 
\end{eqnarray}
} where the superscript $b$ refers to the beam-type, $i = \mu,\bar{e}
(\bar{\mu}, e)$ denotes the neutrino type from $\mu^-(\mu^+)$ beam.
Here $K=N_t n_y$, $N_t$ is number of target nucleons (assumed
isoscalar), $n_y$ is number of years of data and $R$ is the Gaussian
energy resolution function of width $\sigma = 0.07E_\mu$. The
subscripts {\small D} and $\tau$ correspond to the production of
direct muons and $\tau$ production and decay into muons, respectively.
The limits of integration and the restriction from an angular
constraint, in case of tau decay are defined in Ref.~\cite{IS}.  The
charge-blind events are obtained by summing over the index $i$ and the
total events for each beam are then obtained by the sum,
${\cal{R}}^{b}_{tot}(E) = \sum_i ({\cal{R}}^{b}_{i,D}(E) +
{\cal{R}}^{b}_{i,\tau}(E))$~.  We assume 90\% reconstruction
efficiency of muons. Muon event rates accumulated over five years are
used to study the sensitivity to the deviation of the mixing angle
$\theta_{23}$ from maximality.

\begin{figure}
\includegraphics[width=0.4\textwidth]{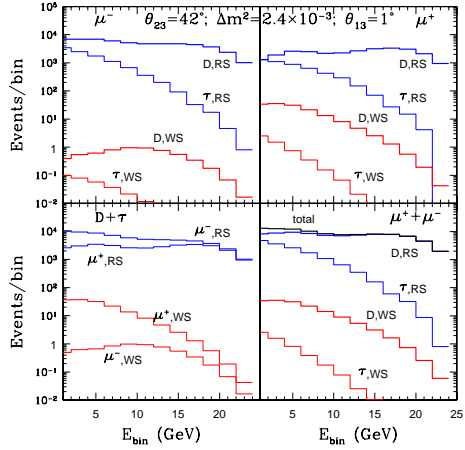}
\caption{Muon event rates as a function of the observed muon energy.
  RS and WS events from $\mu^-$ and $\mu^+$ beams are shown in the
  upper panels. Contributions from direct muon production (denoted by
  $D$) and that of muons from tau decay (labeled as $\tau$) are
  separately shown. The left lower panel shows the sum $D+\tau$.
  Oscillation parameters are as given in text.}
\label{fig:data}
\end{figure}

Typical event rates at L=7400 Km (magic baseline) for oscillation
parameters, $\Delta m^2=2.4\times 10^{-3}$ eV$^2$,
$\theta_{23}=42^\circ$, $\theta_{13}=1^\circ$, $\sin^2 \theta_{12}=
0.304$ and $\Delta_{21}=7.65\times 10^{-5}$ eV$^2$, are shown as a
function of the observed lepton energy in Fig.~\ref{fig:data}. The
panels show the direct muon production and tau decay contributions to
the RS and WS muon events from $\mu^-$ and $\mu^+$ beams.  Since the
neutrino-nucleon cross-sections are larger than those for
anti-neutrinos, the $\mu^-$ ($\mu^+$) beams are preferred for studying
the RS (WS) events where the cross-sections are larger. However, the
full sample without charge identification is preferable, due to higher
detection efficiency. We therefore add the events (RS+WS) from both
$\mu^-$ and $\mu^+$ beams.

It can be seen that there is a substantial contribution to the RS
events from tau decay into muons. The tau contribution alters the
sensitivity to the oscillation parameters; we next discuss whether
these events can be removed through suitable cuts.
\subsection{Cuts on tau contribution}

\begin{figure}
\includegraphics[width=0.4\textwidth]{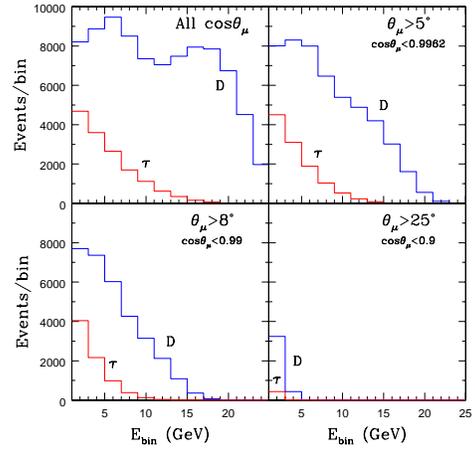}
\caption{Effects of angular cuts on the tau contribution to muon events
at neutrino factories. For details, see the text.}
\label{fig:cuts}
\end{figure}

Since tau production in neutrino-nucleon interactions is extremely
forward-peaked, one obvious way to remove the tau contribution is with
an angular cut. Also, since the tau contribution is substantial at
small observed muon energies where the tau decay rate is large, a muon
energy cut can also be contemplated. The effect of cuts on the event
rates is seen in Fig.~\ref{fig:cuts} -- the only cut effective in
removing the tau contribution is one ($\theta_\mu > 25^\circ$) that
removes the signal itself!  Alternately a muon energy cut of $E >
10-15$ GeV can substantially remove the tau contribution, still
leaving sufficient direct muons. However, such a large energy cut will
worsen the measured precision of the mixing parameters as sensitivity
is higher in the lower energy bins where matter effects are large. In
short, it is not feasible to cut out the tau contribution and still
make a precision measurement, in this case, of the deviation of
$\theta_{23}$ from maximality.

\subsection{Effect of the tau contribution}

As stated earlier, the tau contribution alters the dependence on the
mixing parameters, altering the precision to which we can determine
them. While the tau events have less sensitivity to $\theta_{23}$, the
rate increases while the direct event rate decreases, as $\theta_{23}$
increases towards maximal $\theta_{23}=\pi/4$. The inclusion
of muons from tau events also alters the uncertainties considerably.
A near detector sensitive to muons, measures the combination of flux
times cross-section of the muons. This also appears in the RS event
rate for direct muon production and is therefore well constrained.
However, for tau production and decay, the RS event rate has the
combination of muon flux and the tau production cross-section. The
heavy tau cross-sections have larger uncertainties, where mass
corrections are large. Furthermore, since these contributions result
from oscillations, no near detector can help reduce the uncertainties.
Hence overall uncertainties are much larger for the tau contribution
than for direct muons.

Hence, in our numerical calculations we use an overall
normalization error of 0.1\% for direct, while a modest 2\% is used
for the total (direct+tau), muon events. We use typical input values
of ($\Delta m^2, \theta_{23}, \theta_{13}$) to estimate how well the
generated ``data'' can be fitted, and calculate the resulting
precision on the parameters. We keep the solar parameters fixed at
their best-fit values of Ref.~\cite{schwetz} and set $\delta_{CP}$ to
zero. The best fits (and regions of confidence levels in parameter
space) are obtained by minimizing the chi-squared with a pull
corresponding to the normalization uncertainties specified.

\begin{figure}
\includegraphics[width=0.4\textwidth,bb=18 178 592 672, clip]{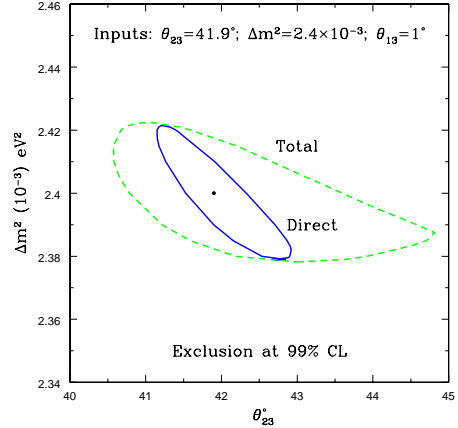}
\caption{Allowed $\Delta m^2$-$\theta_{23}$ parameter space at 99\% CL
  from CC muon events, directly produced (solid line) and with
  inclusion of those from tau decay (dashed line). See text for values
  of input parameters.}
\label{fig:contt13fix}
\end{figure}

\begin{figure}
\includegraphics[width=0.4\textwidth]{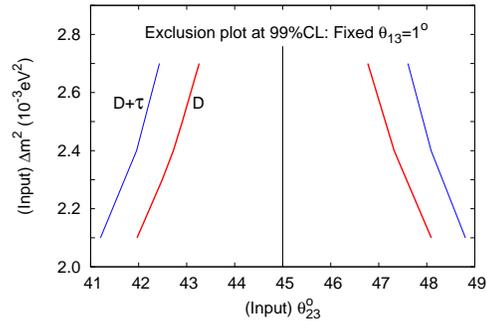}
\caption{The largest (smallest) true value of
$\theta_{23}$ in the first (second) octant that can be discriminated
from $\theta_{23} = \pi/4$, as a function of $\Delta
m^2$ are shown when only Direct ($D$) and total ($D + \tau$) events
are considered; Here $\theta_{13}$ is fixed at $1^\circ$.}
\label{fig:disct13fix}
\end{figure}

We present results for a typical sample set of input parameters,
($\Delta m^2, \theta_{23}, \theta_{13} = 2.4\times 10^{-3}$ eV$^2$,
$41.9^\circ$, $1^\circ$). We minimize first over the pull, and then
over $\Delta m^2$ and $\theta_{23}$, keeping $\theta_{13}$ fixed.
Fig.~\ref{fig:contt13fix} shows the allowed $\Delta
m^2$--$\theta_{23}$ parameter space at 99\% CL. The solid and dashed
lines correspond to considering direct and total (including those from
tau decay) muon events respectively. Note that the 99\% CL contour is
much more constrained with direct than for total muons. In
particular, it is the $\Delta m^2$ values that are smaller than the
input value, that broaden the contour and limit the discrimination.
The largest true value of $\theta_{23}$ that can be discriminated from
maximal is shown in Fig.~\ref{fig:disct13fix}, as a function of
$\Delta m^2$ again, for $\theta_{13} = 1^\circ$. It is seen that tau
contamination worsens the ability to discriminate $\theta_{23}$ from
maximal, thus making this measurement harder than originally expected.

\section{Conclusion}
The oscillations of the muon or electron neutrinos (anti-neutrinos) to
tau neutrinos (anti-neutrinos) results in tau leptons produced through
CC interactions in the detector which on leptonic decay add to the
right as well as wrong sign muon events obtained directly.  This tau
contamination worsens the ability to discriminate against maximal
$\nu_\mu \leftrightarrow \nu_\tau$ mixing. It is practically
impossible to devise satisfactory cuts to remove this tau
contamination. Uncertainties from this tau background must be brought
under control before making precision parameter measurements at
neutrino factories.


%


\begin{theacknowledgments}
NS thanks the organizers for a wonderful conference as well as for
local hospitality.
\end{theacknowledgments}


\end{document}